\documentclass[reprint]{revtex4-1}

\usepackage{hyperref}
\usepackage{epsfig}
\usepackage{graphicx}
\usepackage{subfigure}
\usepackage{latexsym}
\usepackage{color}
\usepackage{fullpage}
\usepackage{epstopdf}
\usepackage{dcolumn}
\usepackage{bm}
\usepackage{ulem}
\usepackage{units}

\newcommand{\major}[1]{\textcolor{black}{#1}}
\newcommand{\minor}[1]{\textcolor{black}{#1}}

\begin{document}

\title{Kinetic pathways to the magnetic charge crystal in artificial dipolar spin ice}

\author{I. A. Chioar$^{1,2}$, B. Canals$^{1,2}$, D. Lacour$^3$, M. Hehn$^3$, B. Santos Burgos$^4$, T. O. Mente\c{s}$^4$, A. Locatelli$^4$, F. Montaigne$^3$, and N. Rougemaille$^{1,2}$}

\affiliation{$^1$ CNRS, Inst NEEL, F-38042 Grenoble, France\\$^2$ Univ. Grenoble Alpes, Inst NEEL, F-38042 Grenoble, France\\$^3$ Institut Jean Lamour, Universit\'{e} de Lorraine and CNRS, F-54506 Vandoeuvre l\`{e}s Nancy, France\\$^4$ Elettra - Sincrotrone Trieste S.C.p.A., S.S: 14 km 163.5 in AREA Science Park, I-34149 Basovizza, Trieste, Italy}

\date{\today}

\begin{abstract}
We investigate experimentally magnetic frustration effects in thermally active artificial kagome spin ice. Starting from a paramagnetic state, the system is cooled down below the Curie temperature of the constituent material. The resulting magnetic configurations show that our arrays are locally brought into the so-called spin ice 2 phase, predicted by at-equilibrium Monte Carlo simulations and characterized by a magnetic charge crystal embedded in a disordered kagome spin lattice. However, by studying our arrays on a larger scale, we find unambiguous signature of an out-of-equilibrium physics. Comparing our findings with numerical simulations, we interpret the efficiency of our thermalization procedure in terms of kinetic pathways that the system follows upon cooling and which drive the arrays into degenerate low-energy manifolds that are hardly accessible otherwise.
\end{abstract}


\maketitle

Artificial spin ice (ASI) are systems designed to explore the intriguing physics observed in magnetically frustrated materials. \minor{Generally fabricated by using lithography techniques}, they offer almost infinite possibilities to construct a wide variety of spin models which can be accessed experimentally in a controlled manner \cite{Nisoli2013}. ASI systems have been the subject of intense research in the last few years and have allowed the investigation of a rich physics and fascinating phenomena, such as the exploration of the extensively degenerate ground state manifolds of spin ice systems \cite{Wang2006, Tanaka2006, Qi2008}, the evidence of new magnetic phases in purely two-dimensional lattices \cite{Moller2009, Chern2011, Rougemaille2011} and the observation of pseudo-excitations involving classical analogues of magnetic monopoles \cite{Ladak2010, Mengotti2010, Rougemaille2013}. Notably, artificial spin ices comprise very different types of systems, including macroscopic arrays of compass needles \cite{Olive1998}, Josephson junctions \cite{Hilgenkamp2003}, superconducting loops \cite{Davidovic1996}, optical traps \cite{Libal2009} and colloidal systems \cite{Yilong2008}.

\begin{figure}
\includegraphics[width=8cm]{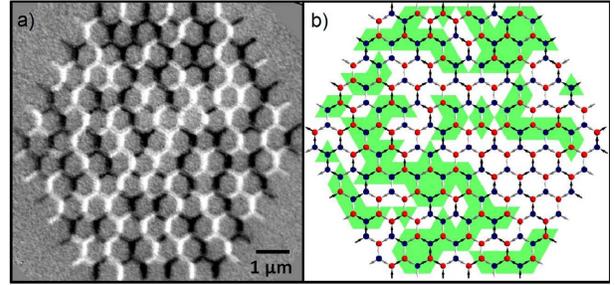}
\caption{\label{fig1}
a) A typical XMCD-PEEM image of our arrays. Black and white contrasts give the local direction of the magnetization of each individual nanomagnet. b) Schematics of the magnetic charge crytallites deduced from a). Red and blue dots correspond to $\pm$1 magnetic charges, respectively. Charge domains are colored in white and green. This image is characterized by a mean \minor{nearest neighbor} charge correlator of $-0.3$.}
\end{figure}

Up until recently, most of the experimental realizations based on magnetic nanostructures were considered as purely athermal systems. Therefore, demagnetization protocols based on the slow decay of an oscillating field are often used to drag such systems into disordered magnetic phases \cite{Wang2007, Ke2008}. However, these protocols show severe limitations in bringing ASI into their predicted low-energy magnetic configurations, where exotic effects emerge, and several other routes have been suggested to make square or kagome ASI thermally active \cite{Morgan2011, Kapaklis2012, Porro2013, Farhan2013-1, Farhan2013-2, Zhang2013, Kapaklis2014, Farhan2014, Montaigne2014}. Up to now, two main directions have been proposed. The first one consists in working close to the blocking temperature of the system, i.e. close to the ferromagnetic / superparamagnetic transition \cite{Farhan2013-1, Farhan2013-2, Kapaklis2014, Farhan2014}. The second approach consists in cooling the system down to the ferromagnetic state starting from a paramagnetic regime obtained above the Curie temperature ($T_{C}$) \cite{Zhang2013, Montaigne2014}.

\begin{figure}
\includegraphics[width=7cm]{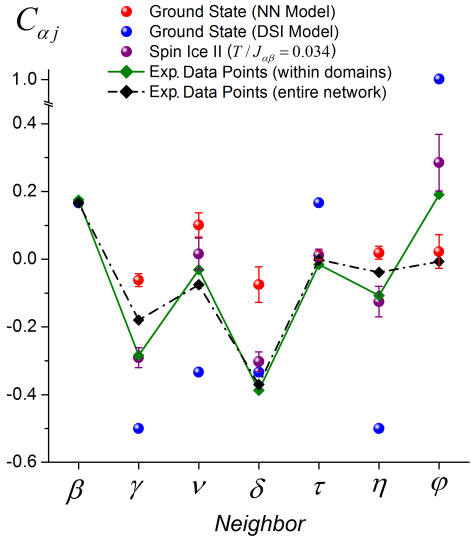}
\caption{\label{fig2}
Average values of the pairwise spin correlators defined up to the seventh neighbor and computed within the charge domains (green diamonds, full line) and at the network-scale (black diamonds, dashed line). For comparison, values expected for the ground state manifolds of the nearest neighbor (NN) SI model and DSI model are represented as red and blue spheres, respectively. The purple spheres correspond to the values given by Monte Carlo simulations for a temperature of $T/J_{\alpha\beta}$ $\simeq 0.034$, which best fits the experimental data. \major{The error bars represent the numerical standard deviations of each correlation type at a given temperature and are computed over the set of sampling Monte Carlo snapshots (see note \cite{Note_Variance})}. The lines linking the experimental data points have no physical meaning and are just guides for the eyes.
}
\end{figure}

Following the second procedure, we show that thermally active artificial kagome arrays of nanomagnets can be locally brought into the magnetic charge crystal phase \cite{Moller2009,Chern2011}. Furthermore, we show that, within these magnetic charge crystallites, pairwise spin correlators are very similar to those expected for the so-called spin ice 2 phase, predicted by Monte Carlo simulations at low temperatures \major{and characterized by a magnetic charge crystal embedded in a partially-ordered kagome spin lattice with preferential spin-loop configurations}. In other words, besides the local observation of alternating $\pm$1 magnetic charges, the spin configurations are also consistent with those specific to the exotic spin ice 2 phase. However, by computing the pairwise spin and charge correlators on the array-scale, we find that the overall magnetic configurations are clearly out-of-equilibrium. Using a kinetic algorithm that \minor{models} how our artificial arrays of nanomagnets behave when crossing the Curie temperature from the paramagnetic state, we manage to reproduce very well our experimental data. We thus interpret our experimental findings in terms of kinetic pathways that the system follows upon cooling \minor{and we further show} how the kinetic process drives the arrays, locally, into low-energy degenerate manifolds that are hardly accessible using a field protocol.

\begin{figure*}
\includegraphics[width=16cm]{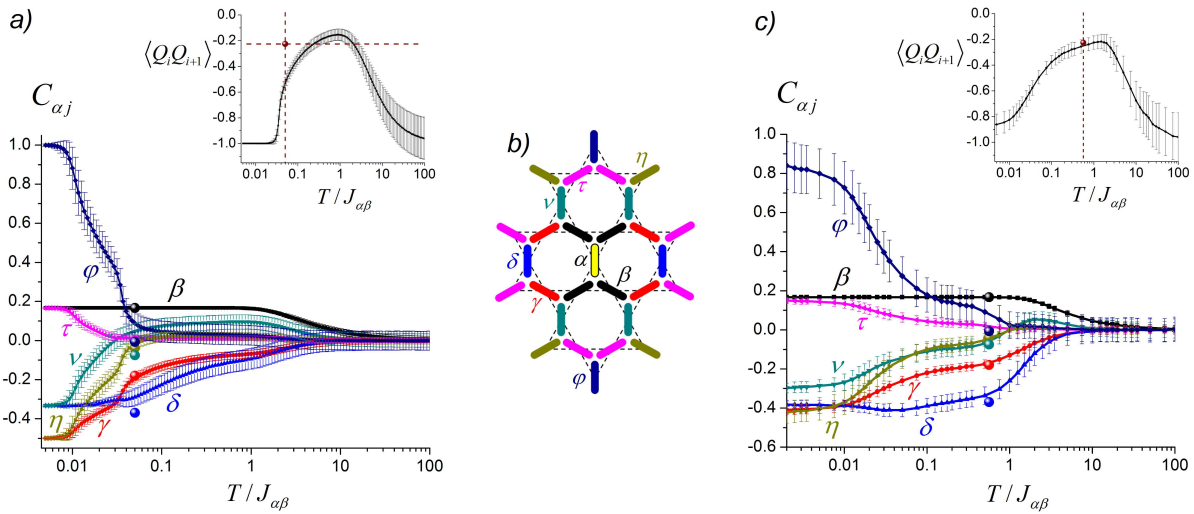}
\caption{\label{fig3}
Pairwise spin and charge (insets) correlators predicted by Monte Carlo (a) and kinetic (c) simulations. In both cases, long range dipolar interactions between spins are taken into account. The colored spheres are the averaged experimental data points computed on the network-scale and further averaged over the 18 different arrays. The definition of the first seven \minor{pairwise spin correlations} (b). The color code is the same for all three images.
}
\end{figure*}

Kagome arrays of 342 nanomagnets were fabricated by e-beam lithography and ion beam etching. Typical dimensions are 500$\times$100$\times$10 nm$^3$ and the nanomagnets are connected at the vertices (see Fig. 1). The constituent material is a ferrimagnetic CoGd alloy which has a Curie temperature adjustable over a wide temperature range by tuning the alloy composition. For the chosen composition (Co$_{0.7}$Gd$_{0.3}$), $T_{C}$ is close to 475 K. \major{Therefore, the networks were annealed at about 500 K, before being cooled down below $T_{C}$ in the absence of the applied magnetic field. Since the remagnetization of the nanomagnets when crossing the Curie point is orders of magnitude faster than the cooling rate (ns and sec time-scales, respectively), the process can be considered as quasistatic}. At room temperature, the nanomagnets are uniformly magnetized, with their magnetization aligned with the long axis of the magnetic elements, and can therefore be considered as Ising pseudo-spins. The resulting magnetic configurations were imaged using X-ray Magnetic Circular Dichroism PhotoEmission Electron Microscope (XMCD-PEEM) at the Nanospectroscopy beamline of the Elettra synchrotron radiation facility \cite{Locatelli2006}. A typical XMCD-PEEM image of a kagome array is shown in Fig. 1a. The magnetic configurations imaged after cooling the system through the Curie temperature show that the arrays are efficiently demagnetized and that the ice rule is strictly obeyed: among more than 3800 observed vertices (corresponding to 18 \minor{different} arrays), no 3-in or 3-out configuration is observed. The magnetic configurations thus fall in the pseudo-spin ice manifold and all vertices are characterized by a $\pm$1 magnetic charge $Q$ \cite{Moller2009}.

From these images, the pairwise spin and charge correlators can be determined \cite{Note_correlators}. First, we have measured the nearest neighbor charge correlator $\langle Q_{i}Q_{i+1} \rangle$ for the 18 arrays we studied. The values range between $-0.30$ and $-0.10$, with an average value of $-0.22$. Although far from the $-1$ value expected for the magnetic charge crystal phase, our measurements are consistent with Ref. \onlinecite{Zhang2013} and indicate that the magnetic charge has crystallized. In fact, we see the formation of charge domains, with a perfect alternation of positive and negative charges on adjacent vertices (see blue/red circles in Fig. 1b). In several cases, these charge domains can extend over a significant fraction of the array and can include more than 1/3 of the total number of vertices. To visualize the distribution of domain sizes, charge domains are colored in white and green in Fig. 1b.

To gain further insight into the underlying physics, we have also considered the pairwise spin correlators, which we computed by averaging within the crystallites of magnetic charges (locally) and over the entire array (globally). \major{As an emergent charge order can be found in both the ground state configuration and the spin ice 2 phase, the local averages of the spin correlations can help discriminate between the two regimes.} Both cases are reported in Fig. 2 up to the seventh neighbor (green and black diamonds, respectively) and each correlation value corresponds to an average performed over the 18 arrays studied. For comparison, the values for the ground state manifolds predicted by the nearest neighbor spin ice (SRSI) model and the dipolar spin ice (DSI) model are also reported (red and blue spheres, respectively) \cite{note_Hamiltonians}. This experimental data shows several remarkable features.

First, the measured pairwise spin correlators can have large values, even for higher order neighbors (see Fig. 2). By no means can these values be obtained with only a nearest neighbor spin-ice model \cite{Wills2002}. Therefore, long-range dipolar interactions cannot be neglected when modeling such arrays.

Second, the spin correlations we measure differ significantly from those of the true ground state of the DSI model. To further determine how far we are from the ground state manifold, we compared our experimental values for the spin correlators with those given by Monte Carlo simulations \cite{Note_Monte_Carlo} by employing a correlation scattering analysis \cite{Chioar2014}. We find that the magnetic configurations we imaged within the charge domains correspond to spin configurations that are very close to the pseudo-ice manifold of the thermodynamic spin-ice 2 phase. In fact, many of our experimental correlations fall within the standard deviations of their corresponding Monte Carlo averages obtained for a temperature of $T/J_{\alpha\beta}$ $\simeq$ 0.034 (see green diamonds and purple spheres in Fig. 2).

Third, if we average the spin correlators computed globally \minor{(see black diamonds in Fig. 2)}, the resulting values significantly differ from those calculated locally, within the charge domains \minor{(green diamonds in Fig. 2)}. In terms of effective temperature, the best fit obtained with our correlation scattering analysis gives a $T/J_{\alpha\beta}$ value of about $0.05$ (note that the $C_{\alpha\nu}$ and $C_{\alpha\delta}$ spin correlators are clearly out of the standard deviations). This $T/J_{\alpha\beta}$ value corresponds to a charge correlator of about $-0.54$ (crossing between the charge correlation plot and the vertical line in the inset of Fig. 3a). This contrasts with the $-0.22$ value that we find experimentally when averaging at a global scale. We then have to solve an apparent contradiction: while pairwise spin correlations measured on the array-scale are those corresponding to configurations approaching the spin ice 2 phase ($T/J_{\alpha\beta} = 0.05$), the charge correlator severely deviates from the at-equilibrium Monte Carlo average found for this temperature (vertical line in the inset of Fig. 3a). In fact, the experimental $\langle Q_{i}Q_{i+1} \rangle$ value alone indicates that the system has barely reached the spin ice 1 manifold (\minor{first} crossing between the charge correlation plot and the horizontal line in the inset of Fig 3.a), although our systems contain large magnetic charge crystallites. As we shall see further on, these differences are signatures of the kinetics of the thermalization process which leads to an out-of-equilibrium state.

To model the experimental procedure, we make the hypothesis that upon cooling below $T_{C}$, each nanomagnet is subject to thermal fluctuations, but once magnetized, magnetization reversal is no longer possible due to the relatively high energy-barriers at stake. Except for the first nanomagnet, the magnetization of a given nanomagnet is biased by the stray field of its environment \cite{Note_Kinetic_Monte_Carlo}.
Therefore, we model the full sample magnetization by the following steps~:
1~- choose randomly a nanomagnet and set the direction of its magnetization,
2 - calculate the resulting stray field over the entire array and pick the not-yet-magnetized nanomagnet that perceives the highest effective field,
3 - set the direction along which this nanomagnet will orient itself, according to a Boltzmann-like probability, 
4 - go back to step 2 and keep repeating steps 2 to 4 until the full sample is magnetized \cite{note2}. Once the array is fully magnetized, we calculate the resulting spin-spin and charge-charge correlators \minor{on the network scale} and repeat these measurements for temperatures ($T/J_{\alpha\beta}$) ranging from $10^{2}$ to $10^{-3}$. The corresponding values of the spin correlators (defined in Fig. 3b) are reported in Fig. 3c with the same color code used for the Monte Carlo simulations.

\begin{figure}
\includegraphics[width=7cm]{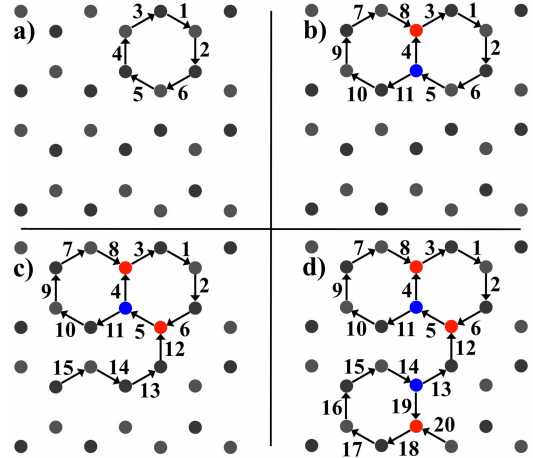}
\caption{\label{fig4}
Snapshots of the kinetic algorithm at a low temperature. The nanomagnets that are magnetic are represented by a black arrow. The numbers indicate the ordering sequence. Red and blue dots correspond to $\pm$1 magnetic charges, respectively, whereas black dots are associated with \minor{charges that have not yet been fully defined}.
}
\end{figure}

\major{Although there are some differences between the correlators deduced from the two numerical approaches, they still show striking similarities. This is mostly due to the fact that the spin interactions in both cases are described by the dipolar spin ice Hamiltonian, hence the energy landscape is the same \cite{note_Hamiltonians}. However, the Monte Carlo approach explores the different configurations at-equilibrium and in an ergodic manner, whereas the kinetic algorithm is a rather one-shot approach, sequentially magnetizing each spin according to a Boltzmann probability in its attempt to minimize the free energy. Therefore, although some correlations can differ in both magnitude and sign for certain temperature windows (see the behavior of the $C_{\alpha\delta}$ and $C_{\alpha\nu}$ correlators for $T/J_{\alpha\beta}$ ranging from $1$ to $0.1$ - Fig. 3), the overall matching is good, especially at low-temperature where our experimental correlations fall.}

By performing a correlation-scattering analysis like in the Monte Carlo case, we can place our experimental values for both the pairwise spin and charge correlators on the temperature-dependent variations predicted by the kinetic model. Interestingly, they all agree upon the same effective temperature, $T/J_{\alpha\beta}$=0.56, and \minor{a vast majority of them} fit within the standard deviations associated to this temperature (see Fig. 3c and inset). The kinetic algorithm thus describes well all our experimental findings and solves the apparent contradiction mentioned above. We emphasize that, if the dipolar interactions are suppressed in the kinetic algorithm, leaving only nearest neighbor interactions at play, the model fails to reproduce our experimental results.

We have finally compared the magnetic configurations we imaged with the ones obtained by applying ac demagnetization protocols, similar to those used in other works \cite{Wang2007, Ke2008, Rougemaille2011}. We find out that, for the same arrays, the effective temperature deduced from the analysis of the pairwise spin correlators is about one order of magnitude lower if the thermal kinetic approach is used. This feature raises further questions on the underlying mechanisms that make one procedure more efficient than the other in driving the system to a low-temperature state and clearly deserves more in-depth analysis. However, we have noticed that the kinetic procedure allows local magnetic configurations that are difficult to obtain through the use of a field protocol. Figure 4 shows snapshots of a kinetic simulation performed at low temperatures and at different steps of the magnetization process. The numbers labeling the nanomagnets indicate in which order they remagnetize. In this temperature regime, the magnetization process favors the formation of full hexagons with flux closure magnetic configurations. These local spin arrangements are specific to low-energy manifolds obtained with at-equilibrium Monte Carlo simulations, in which loop-flip algorithms are used to overcome the critical slowing down behavior encountered by single spin flip protocols \cite{Moller2009, Rougemaille2011}. \major{Interestingly, when working at the superparamagnetic limit, single spin flips are able to drive building blocks of such artificial systems into their corresponding ground state configurations, but they quickly become inefficient as the system size increases \cite{Farhan2014}}. However, the kinetic process at work in this study spontaneously favors closed-loop configurations. We thus relate the efficiency of our thermalization procedure to the kinetic pathways that the system follows when crossing the Curie temperature of the constituent material.

In conclusion, we have studied the behavior of thermally active artificial kagome spin ice systems by cooling the sample from a high-temperature paramagnetic state down below the Curie point of the constituent material. The resulting magnetic configurations present large magnetic charge crystallites and a detailed correlation analysis shows that our arrays are brought, locally, into the spin ice 2 phase, characterized by the emergence of a magnetic charge order within a disordered spin network. The kinetic processes that take place during the cooling procedure appear as an efficient mean to drive the system into low-energy manifolds where exotic physics emerges, opening new avenues to investigate unconventional magnetism in artificial spin systems.

This work was partially supported by the Region Lorraine and the Agence Nationale de la Recherche through the project ANR‐12-BS04-009 'Frustrated'. I.A Chioar acknowledges financial support from the Laboratoire d'Excellence LANEF Grenoble.


\begin{thebibliography}{2}
\bibitem{Note_correlators}
Like in other works, we define the correlation coefficients between spins $\vec{S}_i $ and $\vec{S}_j$ as $C_{ij} = \langle \vec{S}_i.\vec{S}_j \rangle$ and the charge correlator as $\langle Q_{i}Q_{i+1} \rangle$ where $Q_i$ is the magnetic charge on the site $i$ deduced from the dumbbell approximation.
\bibitem{note_Hamiltonians}
\major{Both models have Heisenberg-like Hamiltonians, $H = - \sum_{(i<j)} J_{i j}.\vec{S}_i . \vec{S}_j$. In the case of the nearest neighbor model, the coupling coefficient is constant and limited to nearest-neighbours only, $H_{SRSI} = -J \sum_{<i<j>} \vec{S}_i . \vec{S}_j$. The long-range coupling constant of the dipolar spin ice model is based on the point-dipole approximation in which the coupling coefficients $J_{i j}$ between two magnetic moments, $\mu \, \vec{S}_i$ and $\mu \, \vec{S}_j$, located at $\vec{r}_i$ and $\vec{r}_j$ respectively, can be expressed as: $J_{i j} = - D \, ( \, \frac{1}{|\vec{r}_{ij}|^{3}} - 3 \frac{( \vec{S}_i . \vec{r}_{ij} ) ( \vec{S}_j . \vec{r}_{ij} )}{|\vec{r}_{ij}|^{5} \, \vec{S}_i . \vec{S}_j}) \, $ with $\vec{r}_{ij} = \vec{r}_i - \vec{r}_j$, and $D = ( \mu_0 / 4 \pi) \mu^2$ the strengh of the dipolar coupling. }
\bibitem{Note_Monte_Carlo}
\major{The Monte Carlo simulations were performed for 12$\times$12$\times$3 lattice sites with periodic boundary conditions. $10^4$ modified Monte Carlo steps (MMCS) are used for thermalization and measurements are computed over $10^4$ MMCS. We note that simulations performed for a network of 342 spins with free boundary conditions yield very similar results to the periodic boundary conditions case. Furthermore, the same ground state configuration is found for this particular finite system size as in the infinite network case.}
\bibitem{Note_Variance}
\major{The theoretical values for the spin and charge correlations reported in Figures 2 and 3 correspond to the average values computed over the set of $10^4$ modified Monte Carlo steps (MMCS) used for sampling for each temperature value. Their corresponding standard deviations are also reported in both figures.}
\bibitem{Note_Kinetic_Monte_Carlo}
\major{By tracking the maximal intensity of the stray field throughout the remagnetization process, this algorithm deterministically chooses the island to be remagnetized at each step. This should be contrasted for instance with standard kinetic monte carlo approaches in which each elementary move is associated to a stochastic choice, as is the time increment in these algorithms. In our model, the only stochastic component is found in the choice of an island's magnetization direction, which is performed according to a Boltzmann distribution.}
\bibitem{note2}
Nevertheless, there are many possible ways to choose the first nanomagnet (step 1) and we can even magnetize several nanomagnets at once within this first step. While the corresponding variations are not reported here, we explored several scenarios. Although differing quantitatively, they show similar qualitative results.
\end{thebibliography}
\end{document}